\documentclass[preprint,12pt]{aastex}
\usepackage{natbib}
\usepackage{graphicx}
\usepackage{latexsym}

\graphicspath{{figs/}}
\usepackage[normalem]{ulem}
\begin{document}

\title{Planetesimals in Debris Disks}

\author{Andrew N.~Youdin} 
\author{George H.~Rieke}

\affil{Steward Observatory, University of Arizona, Tucson, AZ, 85721\\
youdin@email.arizona.edu, grieke@as.arizona.edu}

\section{Introduction}

Planetesimals form in gas-rich protoplanetary disks around young stars.   
However, protoplanetary disks fade in about 10 Myr.  The planetesimals (and also many of the planets) left behind are too dim to study directly. Fortunately, collisions between planetesimals produce dusty debris disks.  These debris disks trace the processes of terrestrial planet formation for 100 Myr and of exo-planetary system evolution out to 10 Gyr.
This chapter begins with a summary of  planetesimal formation (see \citealp{houches10, cy10, johansen15} for more detailed reviews) as a prelude to the epoch of planetesimal destruction.  Our review of debris disks covers the key issues, including dust production and dynamics, needed to understand the observations.  Our discussion of extrasolar debris keeps an eye on similarities to and differences from Solar System dust.

\section{The Formation of Planetesimals}
\label{sec:ptmlform}

The first step in the accretion of terrestrial planets and gas giant cores is  the growth of dust grains into planetesimals, typically defined as solids exceeding a kilometer in size.  Models of planetesimal formation have three primary considerations. First, aerodynamic interactions with the gas disk guide the motions of planetesimal building blocks: dust grains, pebbles and boulders.\footnote{Radiative forces, important in debris disks (\S{\ref{sec:dust}}), are less significant in optically thick protoplanetary disks, where the radiation field is nearly isotropic and gas drag is stronger.}  Second, particle sticking and related collisional evolution dominates the early stages of grain growth.  Third, the final assembly of planetesimals likely involves particle concentration mechanisms, which ultimately trigger gravitational collapse.  

\begin{figure}  
\begin{center}  
\includegraphics[width=6.0in]{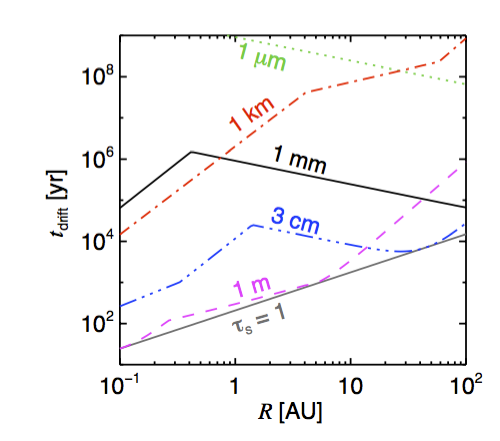}  
\caption{The inward radial drift of solids in a gas disk leads to the ``meter-size barrier" in planetesimal formation. The drift timescale ($t_{\rm drift}$, the orbital radius over inflow speed)  is plotted against orbital distance in the ``minimum mass" disk model of \citet{cy10}.  Curves for various particle sizes are labelled,  with kinks at transitions between drag laws.  Optimal coupling, the gray $\tau_{\rm s} = 1$ curve, gives the fastest radial drift. 
} \label{fig:drift}  
\end{center}  
\end{figure}

Within a protoplanetary disk, particles drift inward because they encounter a headwind of disk gas, which extracts orbital angular momentum.  The headwind arises because gas gets support from radial pressure gradients, an outward acceleration on average.  A meter-sized solid  falls towards its star on timescales of order 100 years. Getting through the infamous ``meter-sized" barrier poses a very stringent constraint on collisional growth \citep{ahn76}.  
 Radial drift is fastest for solids that are optimally coupled to the gas, meaning the dimensionless drag constant $\tau_s = \Omega t_{\rm stop}$ is unity.  The stopping time, $t_{\rm stop}$, is the e-folding timescale for gas drag to damp particle motion, and $\Omega$ is the orbital frequency.   Larger particles have larger $\tau_{\rm s}$, but the radial location and disk model also matter.  Figure \ref{fig:drift} shows that the fastest drifting, $\tau_{\rm s} = 1$ particles are ${\sim}1$m around 1 AU and only ${\sim}3$cm near the Kuiper belt location.

Collisional growth of dust grains up to millimeter sizes is observed in  protoplanetary disks \citep{williams11}.  Small bodies stick due to electrostatic interactions, while gravitational attraction becomes important for massive planetesimals.   Collisional physics is complex, so laboratory experiments must demonstrate which collision types result in growth versus bouncing, erosion or fragmentation \citep{bw08, beitz11}.  Experiments suggest a mm-sized barrier to growth for silicates \citep{zsom10}, although pathways to cm-size and larger bodies by coagulation may exist, especially if ices are present 
\citep[e.g.][]{drc-azkowska14}.  Because cm---m sized solids have low sticking efficiencies and high drift speeds, collisional growth alone may not explain planetesimals.

The gravitational collapse of a sea of small bodies into massive planetesimals offers a top-down alternative to the bottom-up process of collisional growth  \citep{gw73,ys02}.   
The collapse of (for instance) many cm-sized rocks into a 50 km rubble pile would bypass the meter-size barriers.  The key obstacle to gravitational collapse is the mixing of solids by gaseous turbulence \citep{stu80, yl07}.
Collapse is difficult in a smooth (non-clumpy) disk of particles.  The aerodynamic concentration of particles into dense clumps helps  seed gravitational collapse.  Many such concentration mechanisms exist.   As an added bonus, collisions in dense particle clumps are gentler and more conducive to coagulation \citep{JohYou12}.

Particles naturally collect in pressure maxima, which are deviations from the general trend of gas pressure decreasing outwards in disks \citep{whi72}.  Inside the pressure maximum, particles are pushed back toward the maximum since the pressure gradient is reversed and particles see a tailwind that reverses the usual inward drift.  In principle, particles of all sizes can drift towards and collect inside pressure maxima.  However, particles near $\tau_s = 1$ will collect in them fastest, and small $\tau_s \ll 1$ solids can better escape by turbulent diffusion \citep{lyra13}.  Many possible sources of pressure maxima exist (see reviews cited above).
For instance at ice-lines, the vapor-to-solid transition can also change the ionization state and gas accretion rate, leading to a pressure maximum \citep{kretke2007}.  Any such appeal to special locations must explain why planetesimal formation appears so pervasive.

\begin{figure}  
\begin{center}  
\includegraphics[width=6.5in]{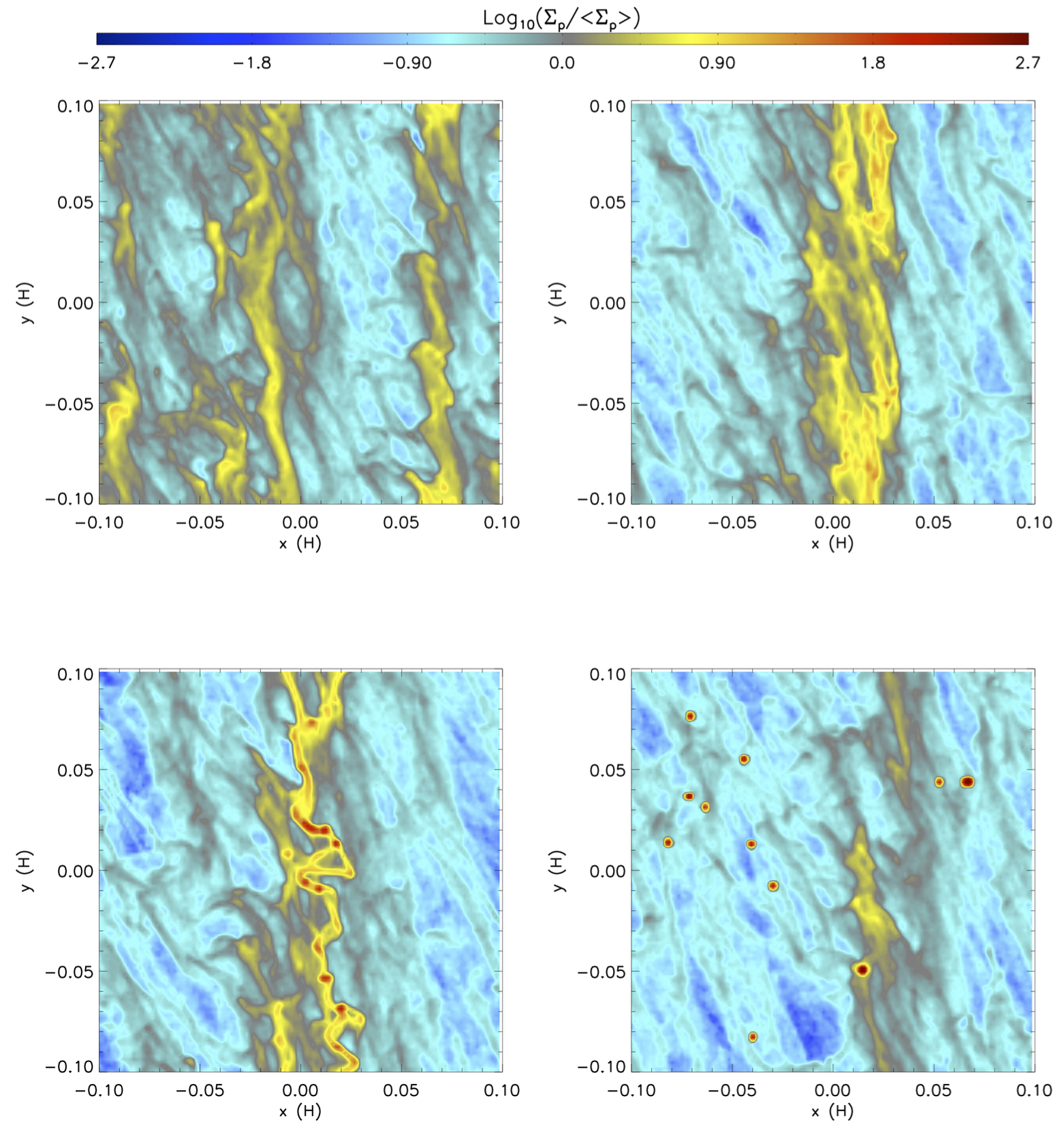}  
\caption{Snapshots from a planetesimal formation simulation \citep{simon15}.  Aerodynamically coupled particles (with $\tau_{\rm s}=0.3$) and gas are simulated in a 3D patch of a disk.  The particle column density is plotted vs. radius and azimuth ($x,y$) in units of the gas scaleheight.  Time evolves from top-left to bottom-right over 10 local orbits.  In the top panels, azimuthally stretched clumps form via streaming instabilities.  In the bottom panels, the clumps fragment into planetesimals with masses equivalent to ${\sim}100$km radii.  
} \label{fig:formation}  
\end{center}  
\end{figure}

The streaming instability is a powerful aerodynamic particle concentration mechanism, which feeds off radial drift and thus requires no special location \citep{yg05,  jy07}.  That drag forces in disks naturally produce particle clumping is related to a traffic-jam effect, but with many subtleties \citep{gp00, yj07, jacquet11}.  For strong clumping, the streaming instability requires particle sizes near optimal coupling,  $\tau_s {\sim} 0.1$--$1.0$, \citep{carrera15} and a mass ratio of particles-to-gas of $\gtrsim 0.02$;  the precise threshold  varies with particle sizes and the  radial pressure gradient \citep{bs10}.  This mass ratio applies to a local patch of the disk (${\sim}10^{-2}$AU wide) which could be enhanced over the global value by radial drift pileups \citep{yc04}.    When these conditions are met, particle concentration is incredibly strong and rapid, readily triggering gravitational collapse, as seen in Figure  \ref{fig:formation}, {\bf  and producing planetesimals with maximum radii around a few hundred km and with} a wide dispersion of sizes \citep{johansen15b}. 

A promising route to planetesimals involves particle coagulation  into sizes large enough for streaming instabilities to produce clumps that gravitationally collapse into planetesimals.  Linking these two mechanisms is a key concern, and gas depletion (increasing $\tau_{\rm s}$ at a given particle size) might be needed in inner disks.  In certain conditions, other particle concentration mechanisms could replace the streaming instability. Small particle concentration via the turbulent cascade is an intriguing possibility  \citep{chs08}, but the ability to produce sufficiently massive planetesimal seeds remains uncertain \citep{pan11}.

\begin{figure}  
\begin{center}  
\includegraphics[width=6.0in]{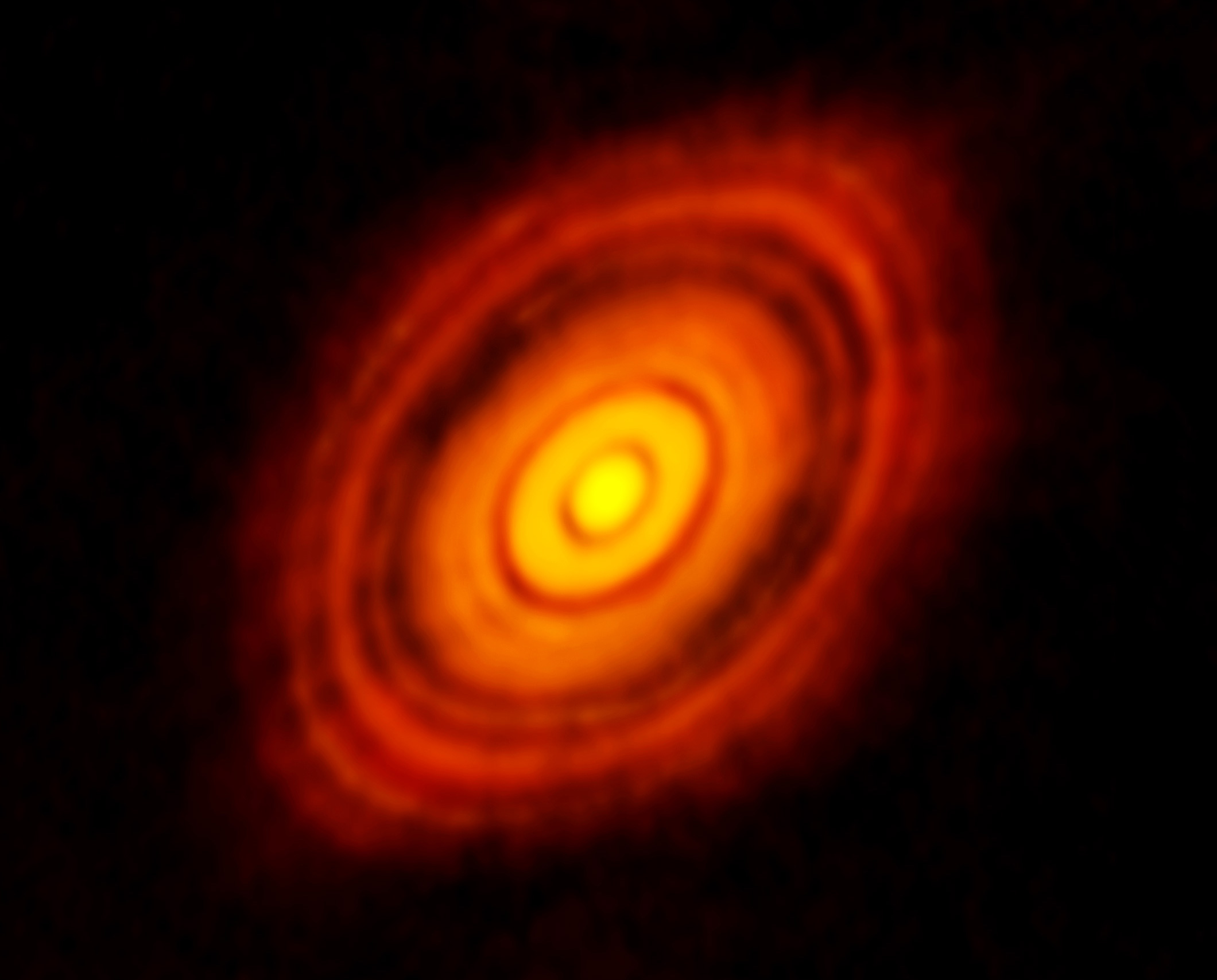}  
\caption{The 240 AU diameter protoplanetary disk around the 1 Myr old star HL Tau \citep{alma-partnership15}. ALMA is sensitive to dust grains with radii  $\sim$100$\mu$m--mm. 
In this image, the central peak is well resolved at roughly 15 AU in diameter. The rings and gaps could be carved by unseen planets or might arise from other dust concentration mechanisms.  Credit: ALMA (NRAO/ESO/NAOJ); C. Brogan, B. Saxton} \label{fig:hltau}  
\end{center}  
\end{figure}

Finally we return to the possibility of direct gravitational instability (GI) from a smooth background.  When drag forces are included, particle GI can occur slowly over long wavelengths, a process known as secular GI \citep{ 
you11a, ShaCuz11}.  More work is needed to understand if secular GI could operate at lower $\tau_{\rm s}$ than streaming instabilities, which would help close any gap between coagulation and dynamical mechanisms.  \cite{takahashi14} predicted that long-lived secular GI could explain dust features such as the (since discovered) rings in the HL Tau system, shown in Figure \ref{fig:hltau}.

However planetesimals  form in protoplanetary disks, many  grow into massive planets, while others do not get that opportunity. 
In the Solar System, remnant planetesimals and minor planets inhabit the asteroid belt, Kuiper belt, Oort cloud and other stable niches. 
 Low planetesimal surface densities and/or dynamical perturbations from neighboring planets prevent such planetesimal belts from growing into larger planets \citep{kenyon12}.
The prevalence of debris disks confirms that both planetesimal formation and leftover belts of remnant planetesimals are common,  as we now describe.

\section{Planetesimals and planetary debris disks}

The first debris disk was accidentally discovered as infrared excess around the supposedly well-behaved calibration star Vega \citep{aumann1984}. Astronomers realized they were seeing dust created in planetesimal belts  tens of AU from the stars \citep{weissman1984, smith1984}.  
A decade later, the first Kuiper Belt object (other than Pluto and Charon, members of the Kuiper Belt independent of planethood!) was found \citep{jewitt1993}, confirming hypotheses that the Solar System had a planetesimal system also at tens of AU from the Sun. This sequence illustrates the importance of debris disks in studying exoplanetary systems,  particularly since space-based infrared telescopes, namely {\it IRAS, ISO, Spitzer, Akari, WISE, and Herschel}, have found hundreds of them.

Planetesimal collisions in debris disks produce copious amounts of  dust through collisional cascades, whereby large planetesimals are gradually ground down to smaller sizes.  Rare giant impacts between protoplanets will  produce a temporary spike in dust production, while a belt of many smaller planetesimals will erode more gradually.  Once dust is ground to micron-sized particles, they are removed by non-gravitational forces.  The need to replenish the dust continually is a key difference between evolved debris disks and younger, gas-rich protoplanetary disks.  Even modest amounts of dust can dominate the surface area of a planetary system.  Dispersing a single 2 km radius planetesimal  into micron-sized grains generates a surface area equal to that of all the Solar System planets.  This dust is readily detectable in 
the infrared when warmed by the star, and in the visible by scattered starlight (using a coronagraph to block direct starlight).

\section{Dust Production and Evolution in Debris Disks}
\label{sec:dust}

\subsection{Collisional Timescales}

Dust production requires frequent collisions. The timescale for planetesimal collisions is
\begin{equation}
t_{\rm c1,2} = {1 \over n_2 \sigma_{1,2} v_{1,2}} \approx {2 \over N_2 \sigma_1 \Omega},
\label{eq:colltime}
\end{equation} 
\noindent
the time for a ``target," with index 1, to  collide with any member of a different-sized ``bullet"  population, with index 2.   The mutual cross section, $\sigma_{1,2}$, is the sum of geometric cross sections in the absence of gravitational focusing, which is weak for high speed destructive impacts \citep[see][for a review of planet formation processes]{YouKen12}. The average relative speed, $v_{1,2}$, is also the impact speed when gravitational focusing is weak.  With $n$ as the volume density, the volumetric collision rate, $n_1/t_{\rm c1,2}$, is naturally symmetric in particle species.  The second, approximate equality contains $N_2$, the vertical column density of bullets, and $\Omega$, the Keplerian orbital frequency.  
This result assumes that: (1)  bullets are smaller than the target so that targets dominate the cross section, $\sigma_1$, while smaller, faster bullets dominate  random motions, $v_2$; and 
(2) compared to eccentricities, orbital inclinations make at least comparable contributions to random motions, giving the scale-height, $H_2 \approx v_2/\Omega$.  Thus $N_2 = 2 H_2 n_2$ gives Equation \ref{eq:colltime}. The removal of uncertain relative speeds makes this result useful.

As an example, consider the Kuiper Belt, with  $\sim0.1 M_\oplus$ of planetesimals between 42--48 AU for a surface mass density $\Sigma \approx 1.6 \times 10^{-3}\;\mathrm{g\; cm^{-2}}$  and a typical internal density $\rho = 2 \; \mathrm{g/cm^3}$\citep{gladman01}. 
We assign a uniform radius, $a = 50$km, to all planetesimals, to overestimate the collision rate between these moderately large KBOs.  The resulting collisional timescale, $t_{\rm c} \approx 800$ Gyr, is incredibly long, confirming that impacts between large KBOs do not currently occur. 

To illustrate that collision rates are dominated by small bullets, we now consider an idealized size distribution that extends to a maximum radius $a_0 = 50$ km as
\begin{equation}
\bar{n}(a) = N_0 {a_0^{q-1} \over a^q} \propto a^{-q}\, ,
\end{equation} 
 with $N_0$  the surface number density used for normalization.  The size distribution $\bar{n}(a) da = dN$ gives the number of bodies (here per unit surface area) in a size bin of width $da$. 
 The cumulative size distribution $N(>a) \sim \int_a^\infty \bar{n}(a)da \propto a^{1-q}$ for $a \ll a_0$.  
We adopt $q = 3$, which gives the mass (dominated by the largest bodies) as $\Sigma \approx (4 \pi/3)\rho N_0 a_0^3$.  The number of objects above a size $a' \ll a_0$ is $N(>a') \approx N_oa_0^2/(2a'^2)$.  Applying Eq.\ \ref{eq:colltime} with $N_2 = N(>a')$ and $\sigma_1 = \pi a_0^2$ gives the collision time  for a large ($a_0$) planetesimals, 
\begin{equation}
t_{\rm c1,2} = {16 \rho a'^2 \over 3 \Sigma a_0 \Omega} \approx 0.7 \left(a' \over 1\; {\rm cm}\right)^2 \;\mathrm{Myr}, 
\end{equation} 
showing the domination by small bullets.  The ability of small bullets to generate dust debris depends on collision speeds and strength laws.

\subsection{Collisional Cascades}

Collisional grinding in debris disks leads to a pseudo-steady-state size distribution, reached in $\sim$10--20 Myr \citep{wyatt2008}.  The shape of the size distribution depends primarily on impact strength (or disruption threshold $Q_D^*$) and secondarily on orbital dynamics.  The classic distribution $\bar{n}(a)\propto a^{-q}$ with $q = 7/2$ \citep{dohnanyi1969} assumes size-independent impact strength and a  steady-state, self-similar solution, i.e.  extending to arbitrarily large and small sizes.  However, $Q_D^*$, the energy per mass required to catastrophically disrupt a target (unbind half the mass), decreases with increasing size for small bodies in the strength-dominated regime.  Above ${\sim}0.1$ km,  disruption is gravity-dominated, and  $Q_D^*$ increases with size. 

A size-dependent $Q_D^*$  affects the predicted $q$  \citep{obrien03}.  In the strength regime, large bodies are weaker and preferentially destroyed, steepening $\bar{n}(a)$ to larger $q$ (and vice-versa in the gravity-dominated regime).
The steady-state approximation is also inexact due to the gradual erosion of the source population.  Combining these effects, \cite{gaspar2012} calculate $q = 3.65$ in the strength-dominated regime,  consistent with observed spectral energy distributions.

A break in the powerlaw $q$ introduces non-self-similar effects, namely waves in the size distribution  \citep{campo-bagatin94, thebault2007}.  For example when small grains are removed, an excess of slightly larger grains develops due to reduced collisional erosion.  However this excess in turn produces a deficit of yet larger bodies.  The wave pattern of alternating excesses and deficits eventually washes out with increasing size, due to the broad range of bullet-to-target mass ratios that can produce debris.

Most collisional size distributions fall between $3{<}q{<}4$ so that larger bodies incorporate most of the total mass while smaller bodies dominate the surface area.  Small $\mu$m-size grains primarily emit  infrared radiation and  also scatter starlight.   Millimeter-wave facilities (like ALMA) are sensitive to larger ($>$100$\mu$m) grains.  The different grain sizes  in the collisional cascade also experience varying non-gravitational forces.

\subsection{Non-gravitational forces on small particles}

The large grains observed at mm and sub-mm wavlengths reside in orbits near the parent bodies that supply the collisional cascade.  Smaller grains are more mobile, due to stronger non-gravitational forces, including
radiation pressure, Poynting-Robertson drag, stellar wind drag, and Lorentz forces \citep{burns1979, gustafson1994}. 

Gravity and radiation pressure forces on dust grains are both inverse square laws, making their dimensionless ratio,

\begin{equation}
\beta = \left |{ F_{\rm r} \over F_{\rm g} }\right|  
=  { 3L_*  Q_{\rm PR}  \over { 16 \pi G M_* c \rho a }  }\, ,
\end{equation}

\noindent
useful, with $Q_{\rm PR}$  the scattering coefficient, $G$ the gravitational constant, $L_*$ the stellar luminosity, $M_*$ its mass, and $\rho$  the density of the grain. Radiation pressure reduces the effective gravitational force by a factor $1-\beta$, and grains with $\beta \ge 1$ experience a repulsive radial force. 

Expulsion occurs when the grain's orbital speed exceeds the modified escape velocity from the system:

\begin{equation}
v_{\rm orb} > v_{esc}=\sqrt {\frac{2GM_\ast}{R}(1- \beta)} \, ,
\end{equation}

\noindent
with $R$  the orbital radius. 
Thus grains with $\beta \ge 0.5$  will pass very quickly out of the system on hyperbolic orbits, assuming grains are released at the  the local circular Keplerian velocity (LCKV). 
Grains  (starting at the LCKV) with $\beta < 0.5$ find themselves on eccentric orbits with periastrons at their release point.  An initial kick that gives a deviation from the LCKV will give a different $v_{\rm orb}$ and critical $\beta$.

From the perspective of an orbiting grain, stellar radiation arrives from a slightly forward-shifted direction due to the aberration of light. Absorbed radiation imparts momentum that  
opposes the orbital motion, producing Poynting-Robertson drag (PRD).  PRD causes inspiral of the grain towards the star and also damps the orbital eccentricity.
Since the  speed of light greatly exceeds orbital velocities,  the aberration angle is small and the force weak:

\begin{equation}
F_{\rm PR} = \frac{v}{c^2}W = \frac{a^2 L_{\rm *}}{4 c^2}\sqrt{\frac{G M_{\rm *}}{R^5}}
\end{equation}

\noindent
where $v$ is the velocity of the grain, $c$ the speed of light, and $W$ the power of the incoming radiation.

PRD is more pronounced for smaller grains closer to their stars. The PRD force weakens with distance as $1/R^{2.5}$ vs.\ $1/R^2$ for gravity.   The gravitational force varies with mass, ${\propto} a^3$, while $F_{\rm PR}{\propto} a^2$.  PRD is thus most significant just above the blow-out size ($\beta{\lesssim} 0.5$) and is already weak for millimeter-size grains.
 The inflow of dust from PRD is insufficient to sustain most observably bright inner disks \citep{wyatt2008}. Instead, this dust must be generated {\it in situ} (e.g., asteroid collisions) or be brought in by comets that disintegrate and deposit the dust.

Stellar wind drag is analogous to PRD, except it is caused by the impact of particles in the stellar wind onto the grain, with an analogous asymmetry in forces in the grain frame of reference. Although usually similar to or 
weaker than PRD, this force can be dominant over PRD for very small grains close to a star \citep{minato2006} or around
young and active late-type stars \citep{plavchan2005}.

Lorentz forces, $F_{\rm Lor}$, are generally negligible. However, near hot stars nanograins can acquire substantial electrical charges and can be retained close to the star to create small near-infrared excesses  from hot ($\sim$1500 K) dust \citep{rieke2015}.

\begin{figure}  
\begin{center}  
\includegraphics[width=6.0in]{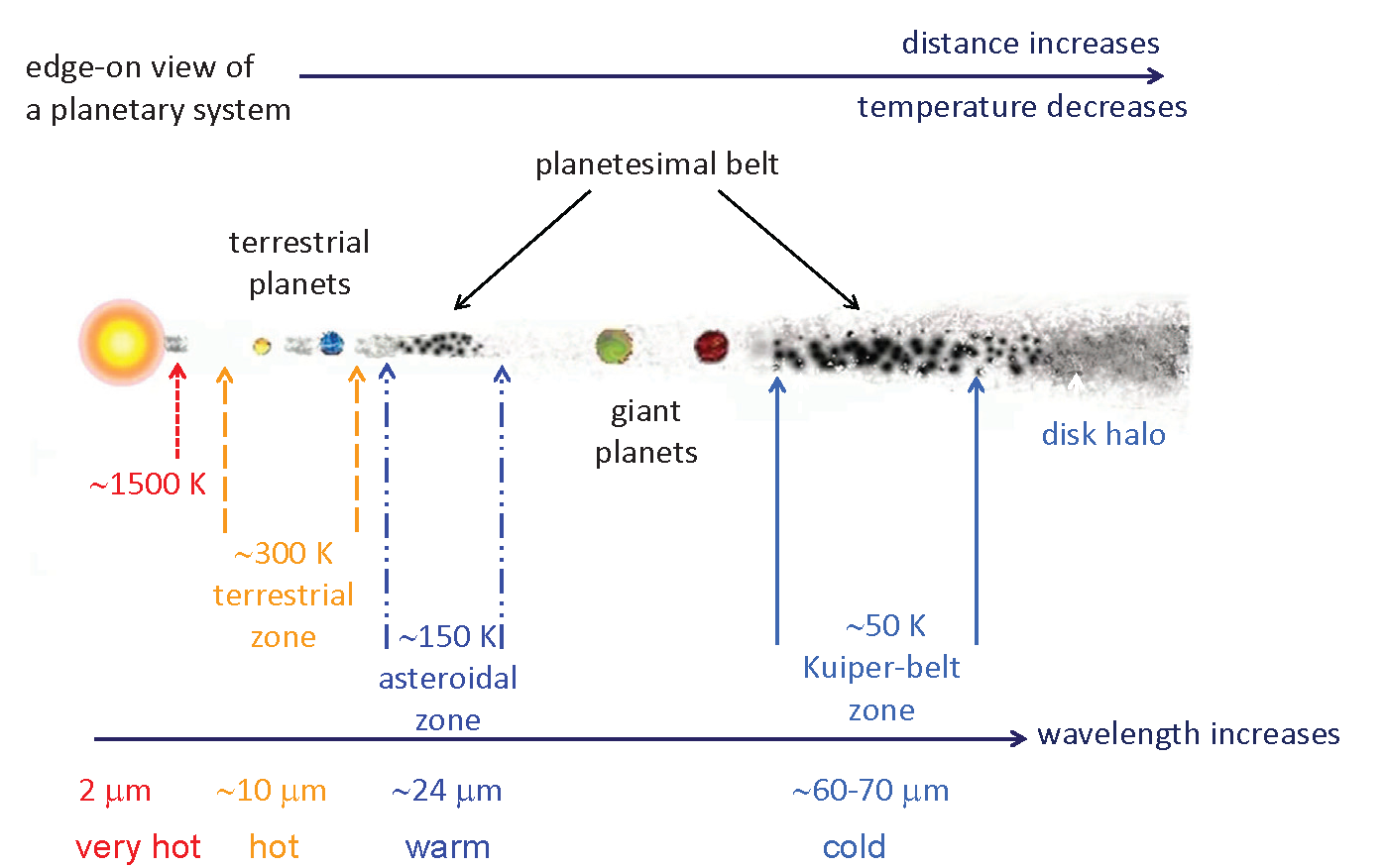}  
\caption{The five characteristic zones of debris disks;  all zones are not detected in every disk \citep{su2014}. The zones are products of temperature, not physical distance from the star, so higher luminosity ($L_{star}$) stars result in scaled-up dimensions, nominally in proportion to $\sqrt{L_{star}}$. For a given $L_{star}$, the nominal dust temperature decreases with distance, $r_d$, as 1/$\sqrt{r_d}$. } \label{fig:debrisstructure}  
\end{center}  
\end{figure}  

\section{Planetesimals and debris: common patterns}

\begin{figure}  
\begin{center}  
\includegraphics[width=6.0in]{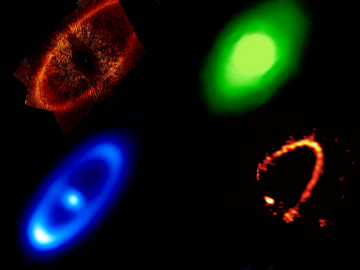}  
\caption{The archetypal debris disk around Fomalhaut. To the lower right, the ALMA image (870$\mu$m) shows the 
parent body ring \citep{boley2012}. At 70$\mu$m (lower left, Herschel; \citealp{acke2012}) the parent body ring is again prominent, 
with the star bright at the center (although the stellar signal is enhanced by a dust component); detailed analysis revelas a halo of $\beta$ meteroids and escaping grains. At 24$\mu$m (upper right, Spitzer;  \citealp{stapelfeldt2004}), the resolution is only slightly poorer than at 70$\mu$m, so structural differences are meaningful. The central peak of the debris system is about 20\% of the total unresolved flux; the remaining 80\% in this image is due to the star. The HST image (upper left; \citealp{kalas2013})  is dominated by small grains efficient at scattering. They line the parent
body belt inner edge and extend beyond out into the halo. [Credit: HST: NASA, ESA, P. Kalas, J. Graham, E. Chiang, E. Kite (University of California, Berkeley), M. Clampin (NASA GSFC), M. Fitzgerald (LLNL), and K. Stapelfeldt and J. Krist (NASA JPL); Spitzer: NASA, JPL-Caltech, K. Stapelfeldt (JPL); Herschel: ESA/Herschel/PACS/Bram Acke, KU Leuven, Belgium; ALMA: A.C. Boley (University of Florida, Sagan Fellow), M.J. Payne, E.B. Ford, M. Shabran (University of Florida), S. Corder (North American ALMA Science Center, National Radio Astronomy Observatory), and W. Dent (ALMA, Chile), NRAO/AUI/NSF] } 
\label{fig:fomalhaut}  
\end{center}  
\end{figure}

We might expect {\it planetesimal} systems to reflect the striking diversity of known {\it exoplanetary} systems. However,  the hundreds of observed debris systems reveal striking similarities.
Figure \ref{fig:debrisstructure}  illustrates the prevalent debris disk zones and Figure \ref{fig:fomalhaut} 
shows an example.

\subsection{\emph{Very hot} dust}

Within a few tenths of an AU of the Sun, only the most refractory grains can survive at temperatures ${\sim}1500$ K.  Moreover, grains in the 0.1 - 1$\mu$m size range are ejected by radiation pressure \citep{gustafson1994}. Nonetheless, a population of $\sim$nanometer-size grains exists, produced by grain-grain collisions and  the sublimation of larger grains. These nanograins acquire electrical charge by electron impact and the photoelectric effect and are trapped in the Solar magnetic field  by Lorentz forces \citep{mann2007}. Some stars (mostly A-type) have much larger amounts of very hot dust lying within $\sim$0.2 AU \citep[e.g.,][]{absil2013}. The spectral energy distribution of this component falls steeply toward longer wavelengths, indicating a lack of grains larger than 200 nm in radius. Again, these grains are probably retained by Lorentz force \citep{rieke2015}.

\subsection{\emph{Hot} dust in the terrestrial planet zone}

\begin{figure}  
\begin{center}  
\includegraphics[width=5.0in]{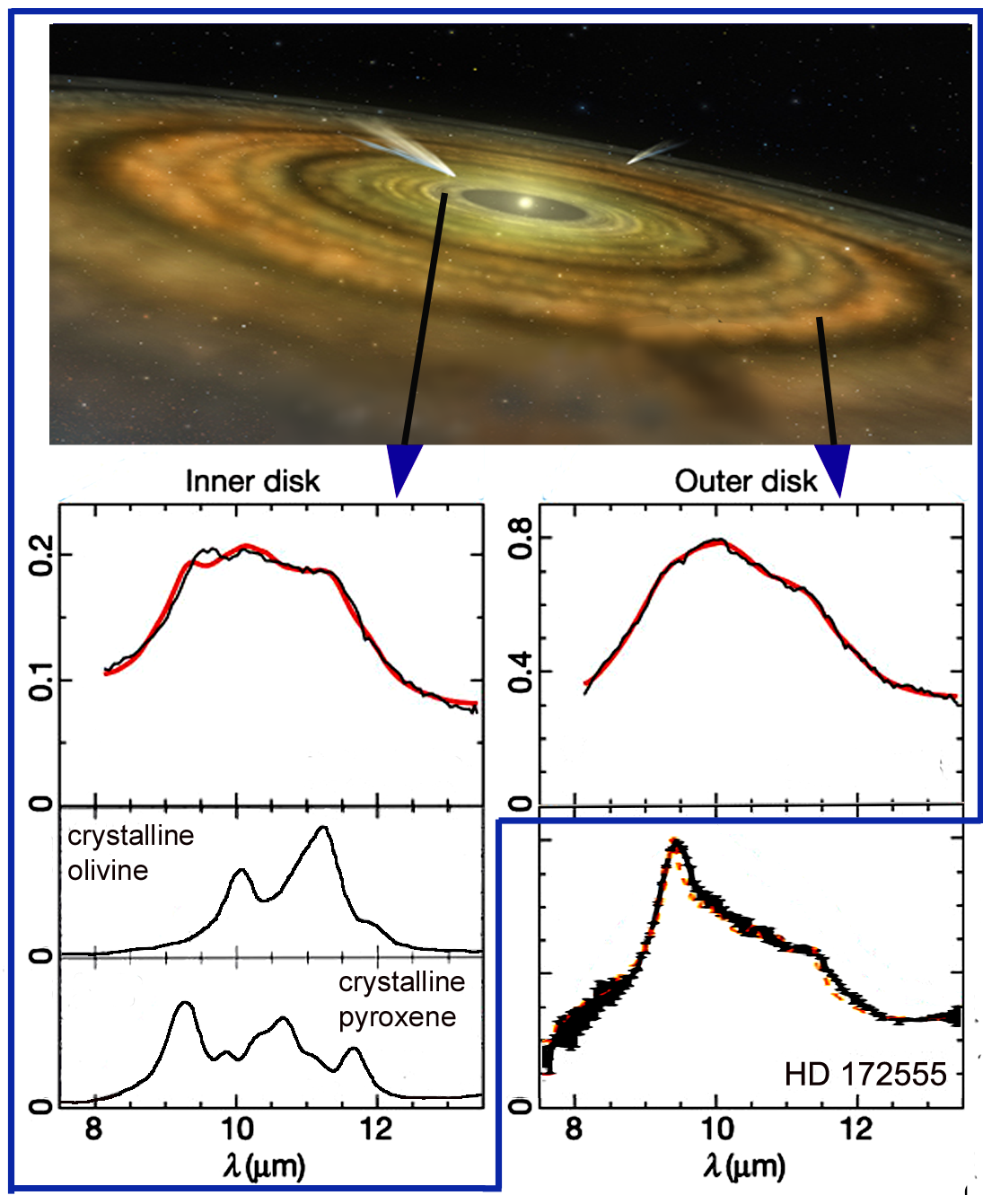}  
\caption{Silicate emission varies with the location in, and evolutionary state of, debris disks. The top image schematically represents a young debris disk. The hot inner disk of HD 163296 (middle left; spectrum in black, model in red)  shows silicate emission.  The model contains olivine and pyroxene silicates in crystalline form, as expected for high temperatures  (lower left); however, grain growth washes out some spectral features. The outer disk has a smooth 10$\mu$m peak (middle right), well fit by unprocessed amorphous silicates. [Credit: NASA/FUSE/Lynette Cook, \citet{vanboekel2004}]. HD 172555 (lower right) shows a sharp 9$\mu$m feature, characteristic of silica grains recently condensed 
from vapor \citep{john2012}.} 
\label{fig:silicates}  
\end{center}  
\end{figure}

Near the Earth, a very dilute population of dust scatters sunlight to yield the zodiacal light and emits primarily in the $10-20$$\mu$m region. The 10$\mu$m silicate emission feature reveals that some of these hot ($\sim$300 K) grains are $\sim$micron-sized\footnote{As the size increases above a few microns, the silicate feature is rapidly reduced because the grain becomes optically thick \citep[e.g.,][]{papoular1983}.} \citep[see Figure \ref{fig:silicates}]{ootsubo2009}. However, the silicate feature is subtle; most of the dust is 10 - 40$\mu$m in size. The grain population is maintained by the disintegration of Jupiter Family Comets and the breakup of asteroids, with the inflow resulting mostly from PRD \citep{nesvorny2010}. 
Hot dust is observed in many  debris disks as subtle silicate emission features \citep{ballering2014, mittal2015}.  
In some cases, the hot dust component has been resolved \citep[e.g.,][]{mennesson2014}; in the 
best-studied examples, the dust is relatively close to the star, inside the zone thermally equivalent to the orbit of the Earth \citep{stock2010, defrere2015}.

\subsection{\emph{Warm} (asteroidal) and \emph{cold} (Kuiper Belt-like) dust}

Going outward from the Sun, the first significant planetesimal/debris disk system is the asteroid belt, with dust produced primarily by the disintegration of Jupiter Family Comets \citep{nesvorny2010}, but with a significant contribution from asteroid collisions.  Indeed, dust bands in the zodiacal light trace the production of collisional asteroid families over the past $\sim$10 Myr \citep{grogan2001, nesvorny2003}. The inflection in the asteroid size distribution near 25 km radius indicates that larger bodies are indestructible, while smaller ones participate in the collisional cascade  \citep[e.g.,][]{obrien2011}. Still, the low volume density of asteroids gives a low fractional luminosity of dust emission, $L_{dust}/L_{star}{\sim}10^{-7}$ \citep{nesvorny2010}, nearly two orders of magnitude below detection limits around other stars \citep{roberge2012}. 

Continuing outward, there is a gap and then the Kuiper Belt debris disk, with a sharp inner edge shepherded by Neptune \citep{liou1999} and resonant structures at larger radii \citep[e.g.,][]{adams2014}. {\it In situ} dust sampling by the Student Dust Counter on the New Horizons mission \citep{szalay2015} confirms that the Kuiper Belt, despite the lower planetesimal surface density and faster orbits,  is a venue for collisional-cascade-based dust production.  The far-infrared emission from  Kuiper Belt dust is estimated to be only about 1\% of the Solar photospheric output at these wavelengths, below current detection limits around other stars \citep{vitense2012}. 

Like the Solar System, debris disks with warm components overwhelmingly come as two-temperature systems \citep{kennedy2014}: a far-infrared component with cold temperatures,  50--100K and a  warm, 150--200 K, component \citep[e.g.,][]{ballering2013, chen2014}.  Large planets could naturally clear the  region between the dust belts \citep[e.g.,][]{su2013}. 
The warm components fade in $\sim$300 Myr \citep{gaspar2013}; the majority of debris disks are detected only as a cold excess. Bright far infrared excesses, in comparison, persist for about 3.5 GYr and are seen roughly independently of stellar type at the 20\% level around stars of late F through early K \citep{sierchio2014}. These outer belts are also accessible to scattered-light imaging, from which we have obtained the highest resolution images of the small dust distributions \citep[e.g.,][]{schneider14, soummer14}.

The similar locations of the Kuiper Belt and outer debris disks suggests that massive planets also reside in debris systems. The situation for the asteroid belt is more complex. It overlaps with plausible locations for the snow line in the Sun's protoplanetary disk \citep{kennedy2008}.  Pressure enhancements at the snow line can provide a favorable site for planetesimal formation, see \S{\ref{sec:ptmlform}}.  But why would such planetesimals fail to assemble into more massive planets, given that the primordial asteroid belt was two to three orders of magnitude more massive than today \citep[][]{morbidelli15}? 
Somehow, an early-formed Jupiter limited planetesimal growth and/or ejected most asteroids. In the Grand Tack scenario, Jupiter may even have migrated inwards through the belt to the location of Mars and then back out \citep{walsh11}, strongly affecting the structure of the belt.  It is thought that the asteroid belt then reformed from bodies inside and outside of its current location \citep{walsh12}.  Alternatively,  warm debris  might indicate the disintegration of huge comet populations, analogous to  Jupiter Family Comets.

\subsection{Dust halos}

Outside the Kuiper Belt there must be a dilute halo of small grains whose motions are strongly affected by radiation pressure  - both $\beta$ meteroids on highly elliptical marginally bound orbits, and also particles escaping into interstellar space.  Similar but far more dramatic halos are detected around a number of prominent debris disks such as $\beta$ Pic \citep{augereau2001} and Fomalhaut (Figure \ref{fig:fomalhaut}), in scattered light as well as thermal emission.

\section{Evolution of planetesimal systems}

Many exoplanetary systems become virtually invisible after their natal protoplanetary disks dissipate, by about 10 Myr of age. The similarities in the Solar System and exoplanetary debris disks embolden us to use debris disks to reveal the development of the invisible exoplanetary systems themselves. Debris disks complement radial velocity and transit studies, which focus on the inner regions of  mature planetary systems. 

The {\it Kepler} mission revealed that planets within only 0.5 AU of G, K and especially M stars outnumber the stars themselves \citep{you11b, mulders15, dressing15}.
In contrast to this bounty, the  mass in protoplanetary disks -- inferred from dust observations -- is not overly generous \citep{williams11}. Significant early coagulation into larger grains avoids an outright conflict with the large incidence of planetary systems \citep{sheehan2014}. 

The most common planets appear to have  masses between Earth and Neptune \citep[e.g.,][] {malhotra2015}, far below  direct detection limits (i.e., imaging). 
Such unseen planets leave traces in debris disks; for example, within the Solar System, the inner edge of the Kuiper Belt is shepherded by Neptune \citep{liou1999}.   Gaps between warm and cold debris belts can also be explained by intervening planets, allowing some understanding of their incidence. 

Debris disks can also trace several phases in planetary system evolution, including:  1.) violent dynamical activity during terrestrial planet growth; 2.)  gradual erosion of remnant belts of planetesimals; and 3.) late phase bursts of activity caused by orbital instabilities among planets, such as the Late Heavy Bombardment in the Solar System.

\subsection{Formation of terrestrial planets}

Terrestrial planets may begin their assembly during the protoplanetary disk phase, with the rapid accretion of pebbles and boulder-sized solids \citep[e.g.,][]{morbidelli15b}.  Aerodynamic drift brings solids from the outer disk, while gas drag also enhances the capture probability of large pebbles \citep{ormel10}.

For traditional planetesimal-dominated accretion, however, the isolation mass plays a key role. It  is reached when protoplanets have accreted most of the planetesimals in their feeding zones \citep{YouKen12}, marking the endpoint of the rapid phase of oligarchic planetesimal accretion \citep{ki98}.    Earth and Venus are much larger than the isolation mass; the merger of isolated oligarchs into such planets requires another 10-100 Myr of intermittent giant impacts \citep{chambers04, kb06, kleine09}, like the Moon-forming impact when the Solar System was 30--100 Myr old \citep[e.g.,][]{jacobson14}. 
Debris disks can trace this process via the high rate of collisional grinding of planetesimals stirred by growing terrestrial protoplanets \citep{kenyon04}  and also by the large amounts of dust produced in individual violent collisions, some of it condensed directly from vaporized rock.

Violent impacts during this era are revealed by the 9$\mu$m silica emission feature, see Figure \ref{fig:silicates}. Silica dust can condense directly from gas produced from silicate minerals in violent collisions \citep{johnson2012};  gaseous silica was a major product of the collision that formed our Moon \citep{canup2004}. HD 172555, a  member of the 20 Myr-old \citep{soderblom2014} $\beta$ Pic moving group, is an outstanding example of a disk with a strong silica feature \citep{lisse2009}. \citet{john2012} conclude that the dust is produced at $\sim$6 AU from the star; given the luminosity of HD 172555, this distance is thermally equivalent to about 2 AU in the Solar System, i.e., just outside the orbit equivalent to that of Mars.  

 \citet{jackson2012} concluded that the aftermath of the Moon-forming collision was a circumsolar ring of debris detectable (with current infrared astronomy technology from a nearby planetary system) for up to 25 Myr. Even greater variability has been seen in the most extreme debris systems \citep{melis2012, meng2012, meng2015}.
For example, the star ID8 in the 35 Myr old NGC 2547 cluster changes its output on monthly timescales  \citep{meng2014}. The debris orbits at $\sim0.33$ AU from a star similar in luminosity to the sun, i.e., near the equivalent position of Mercury. Both HD 172555 and ID8 are estimated to have about $1 \times 10^{19}$ kg of finely divided dust, indicating asteroids at least 200 km in diameter must have been destroyed around them. 
More evidence for violent events in the era of terrestrial planet formation comes from ALMA studies of $\beta$ Pic  in the J = 3-2 transition of $^{12} $CO \citep{dent2014}.  The massive gas cloud on (only) one side of the disk could arise either from the outward migration of a planet that locally enhances planetesimal collisions, or from debris of a single recent planetesimal collision.

\subsection{Long term dust production}

After the mayhem of planet formation has passed, debris disks continue to produce lower levels of dust.  As the population of destructible planetesimals is eroded, dust production from collisional cascades decays  with time \citep[e.g.,][]{wyatt2007}.
 Since collision rates are proportional to the instantaneous mass of the debris disk, this  fading  is nearly independent of the disk mass; more massive disks simply undergo a brief initial phase of rapid collisional destruction. The time evolution of debris disk emission is consistent with this premise  \citep{gaspar2013}, assuming a distribution of initial disk masses proportional to that of protoplanetary disks. This behavior provides a simple baseline against which to identify deviations.  

\subsection{Late stage dynamical shakeups}

Anomalously large excesses likely reflect a dynamical shakeup of the planetary system and enhanced gravitational stirring of its planetesimals. There is a residual (a few \%) of disks detected at 24$\mu$m well beyond the age of $\sim$600 Myr when the baseline would predict no more activity \citep{gaspar2013}. Many of these systems show features in their mid-infrared spectra indicating very small dust grains with short lifetimes around their stars, indicating recent collisions producing enhanced levels of dust. Breaking up a Ceres-sized  planetesimal should yield a detectable excess at 24$\mu$m for about 1 million years \citep{kral2015}.

More extensive dynamical activity has  more profound effects. \citet{booth2009} simulated the effect of the  Late Heavy Bombardment (in the context of the Nice model) on the history of the Solar System's debris disk. The 24$\mu$m emission spikes by a factor of $\sim$3 , but rapidly fades below detection limits in $\sim$30 Myr.  The 70$\mu$m emission does not spike but remains detectable much longer, $>$300 Myr.

\section{Conclusion}

Planetesimals are the building blocks of planet systems. We are developing an understanding of how they are formed and grow in their first 10 Myr, although substantial questions remain such as: 1.) how far past mm-sizes does collisional growth proceed (at different locations) during the lifetime of protoplanetary disks; and 2.)  what combination of dynamical processes (e.g. streaming instabilities, pressure traps and gravitational instabilities) complete the growth of planetesimals if/when standard collisional growth stalls.  Past this era, we can trace planetesimal evolution through debris disks, from which we have found:

\begin{itemize}

\item
There is a remarkable similarity in the positions of the planetesimal belts in the Solar System and around other stars.

\item
This layout probably reflects the influence of unseen giant planets in the exoplanetary systems.

\item
Debris disks exhibit the processes in the building of terrestrial planets through a.) spectral features indicative of the generation of silica dust in violent collisions; and b.) variability in dust production reflecting ongoing planetesimal collisions.

\item
Although most debris disks fade away in accordance with theoretical expectations, a few percent are found around older stars and indicate late phases of dynamical activity in their planet systems.

\end{itemize}


\begin{thebibliography}{}




\bibitem[Absil et al.(2013)]{absil2013}
Absil, O., Defr\`ere, D, Coude\`e du Foresto, V., Di Folco, E. et al. 2013. A near-infrared interferometric survey of debris-disc stars. III. First statistics based on 42 stars observed with CHARA/FLUOR. {\it A\&A}, 555, 104.

\bibitem[Acke et al.(2012)]{acke2012}
Acke, B., Min, M., Dominik, C. et al. 2012. Herschel images of Fomalhaut. An extrasolar Kuiper belt at the height of its dynamical activity. {\it A\&A}, 540, 125.

\bibitem[Adachi et al.(1976)]{ahn76}
{{Adachi}, I., {Hayashi}, C., \& {Nakazawa}, K.} 1976. The gas drag effect on the elliptical motion of a solid body in the primordial solar nebula. {\it  Prog. Theor. Phys.}, 56, 1756.

\bibitem[Adams et al.(2014)]{adams2014}
Adams, E. R., Gulbis, A. A. S., Elliot, J. L. et al. 2014. De-biased populations of Kuiper Belt Objects from the Deep Ecliptic Survey. {\it AJ}, 148, 55.


\bibitem[ALMA Partnership(2015)]{alma-partnership15}
{{ALMA Partnership}, {Brogan}, C.L., {Perez}, L.M. et al.} 2015. The 2014 ALMA long baseline campaign: first results from high angular resolution observations toward the HL Tau region. {\it ApJL}, 808, 3.


\bibitem[Augereau et al.(2001)]{augereau2001}
Augereau, J. C., Nelson, R. P., Lagrange, A. M. et al. 2001. Dynamical modeling of large scale asymmetries in the beta Pictoris dust disk. {\it A\&A}, 370, 447.

\bibitem[Aumann et al.(1984)]{aumann1984}
Aumann, H. H., Beichman, C. A., Gillett, F. C. et al. 1984. Discovery of a shell around Alpha Lyrae, {\it ApJL}, 278, 23.

\bibitem[Bai \& Stone(2010)]{bs10}
{{Bai}, {X.-N.} and {Stone}, J. M.} 2010. The effect of the radial pressure gradient in protoplanetary disks on planetesimal formation. {\it ApJL}, 722, 220.

\bibitem[Ballering et al.(2013)]{ballering2013}
Ballering, N. P., Rieke, G. H., Su, K. Y. L., \& Montiel, E. 2013. A trend between cold debris disk temperature and stellar type: implications for the formation and evolution of wide-orbit planets. {\it ApJ}, 775, 55.

\bibitem[Ballering et al.(2014)]{ballering2014}
Ballering, N. P., Rieke, G. H., \& G\'asp\'ar, A. 2014. Probing the terrestrial regions of planetary systems: warm debris disks with emission features. {\it ApJ}, 793, 57.

\bibitem[Beitz et al.(2011)]{beitz11}
{{Beitz}, E., {G{\"u}ttler}, C., {Blum}, J., {Meisner}, T., {Teiser}, J., \& {Wurm}, G.} 2011. Low-velocity collisions of centimeter-sized dust aggregates. {\it \apj}, 736, 34.



\bibitem[Blum \& Wurm(2008)]{bw08}
{{Blum}, J. and {Wurm}, G.} 2008. The growth mechanisms of macroscopic bodies in protoplanetary disks. {\it \araa}, 46, 21.

\bibitem[Boley et al.(2012)]{boley2012}
Boley, A. C., Payne, M. J., Corder, S. et al. 2012. Constraining the planetary system of Fomalhaut using high-resolution ALMA observations. {\it ApJL}, 750, 21.

\bibitem[Booth et al.(2009)]{booth2009}
Booth, M., Wyatt, M. C., Morbidelli, A., Moro-Mart\'in, A., \& Levison, H. F. 2009. The history of the Solar system's debris disc: observable properties of the Kuiper belt. {\it MNRAS}, 399, 385.

\bibitem[Burns et al.(1979)]{burns1979}
Burns, J. A., Lamy, P. L., \& Soter, S. 1979. Radiation forces on small particles in the solar system. {\it Icarus}, 40, 1.

\bibitem[Campo Bagatin {et~al.}(1994)]{campo-bagatin94}
{Campo Bagatin}, A., {Cellino}, A., {Davis}, D. R., {Farinella}, P., \&
  {Paolicchi}, P. 1994. Wavy size distributions for collisional systems with a small-size cutoff. {\it \planss}, 42, 1079.

\bibitem[Canup(2004)]{canup2004}
Canup, R. M. 2004. Simulations of a late lunar-forming impact. {\it Icarus}, 168, 433


\bibitem[Carrera et al.(2015)]{carrera15}
{{Carrera}, D., {Johansen}, A., \& {Davies}, M. B.} 2015. How to form planetesimals from mm-sized chondrules and chondrule aggregates. {\it A\&A}, 579A, 43.

\bibitem[Chambers(2004)]{chambers04}
{{Chambers}, J. E.} 2004. Planetary accretion in the inner Solar System. {\it Earth Plan. Sci. Let.}, 223, 241.



\bibitem[Chen et al.(2014)]{chen2014}
Chen, C. H., Mittal, T., Kuchner, M. et al. 2014. The Spitzer Infrared Spectrograph Debris Disk Catalog. I. Continuum analysis of unresolved targets. {\it ApJS}, 211, 25.

\bibitem[Chiang \& Youdin(2010)]{cy10}
{{Chiang}, E. \& {Youdin}, A. N.} 2010. Forming planetesimals in Solar and extrasolar nebulae. {\it Annual Review of Earth and Planetary Sciences}, 38, 93.






\bibitem[Cuzzi et al.(2008)]{chs08}
{{Cuzzi}, J. N., {Hogan}, R. C., \& {Shariff}, K.} 2008. Toward planetesimals: dense chondrule clumps in the protoplanetary nebula. {\it \apj}, 687, 1432.


\bibitem[Defr\`ere et al.(2015)]{defrere2015}
Defr\`ere, D., Hinz, P. M., Skemer, A. J. et al. 2015. First-light LBT nulling interferometric observations: warm exozodiacal dust resolved within a few AU of $\eta$ Crv. {\it ApJ}, 799, 42.

\bibitem[Dent et al.(2014)]{dent2014}
Dent, W. R. F., Wyatt, M. C., Roberge, A. et al. 2014. Molecular gas clumps from the destruction of icy bodies in the $\beta$ Pictoris debris disk. {\it Science}, 343, 1490.


\bibitem[Dohnanyi(1969)]{dohnanyi1969}
Dohnanyi, J. S. 1969. Collisional model of asteroids and their debris. {\it J. Geophys. Res.}, 74, 2531.


\bibitem[Dr{\c a}{\.z}kowska et al.(2014)]{drc-azkowska14}
{{Dr{\c a}{\.z}kowska}, J., {Windmark}, F., \& {Dullemond}, C. P.} 2014. Modeling dust growth in protoplanetary disks: The breakthrough case. {\it \aap}, 567, A38.

\bibitem[{{Dressing} \& {Charbonneau}(2015)}]{dressing15}
{Dressing}, C. D., \& {Charbonneau}, D. 2015. The occurrence of potentially habitable planets orbiting M Dwarfs estimated from the full Kepler dataset and an empirical measurement of the detection sensitivity. {\it \apj}, 807, 45.







\bibitem[G\'asp\'ar et al.(2012)]{gaspar2012}
G\'asp\'ar, A., Psaltis, D., Rieke, G. H, \& \"Ozel, F. 2012. Modeling collisional cascades in debris disks: steep dust-size distributions. {\it ApJ}, 754, 74.

\bibitem[G\'asp\'ar et al.(2013)]{gaspar2013}
G\'asp\'ar, A., Rieke, G. H., \& Balog, Z. 2013. The collisional evolution of debris disks. {\it ApJ}, 768, 25.

\bibitem[{{Gladman} {et~al.}(2001){Gladman}, {Kavelaars}, {Petit},
  {Morbidelli}, {Holman}, \& {Loredo}}]{gladman01}
{Gladman}, B., {Kavelaars}, J. J., {Petit}, J.-M. et al. 2001. The structure of the Kuiper Belt: size distribution and radial extent. {\it \aj}, 122, 1051.

\bibitem[Goldreich \& Ward(1973)]{gw73}
{{Goldreich}, P. \& {Ward}, W. R.} 1973. The formation of planetesimals. {\it \apj}, 183, 1051.

\bibitem[Goodman \& Pindor(2000)]{gp00}
{{Goodman}, J. and {Pindor}, B.} 2000. Secular instability and planetesimal formation in the dust layer. {\it Icarus}, 148, 537.


\bibitem[Grogan et al.(2001)]{grogan2001}
Grogan, K., Dermott, S. F., \& Durda, D. D. 2001. The size-frequency distribution of the Zodiacal Cloud: evidence from the Solar System dust bands. {\it Icarus}, 152, 251.

\bibitem[Gustafson(1994)]{gustafson1994}
Gustafson, B.\AA.S. 1994. Physics of Zodiacal dust. {\it AREPS}, 22, 553.




\bibitem[Jackson \& Wyatt(2012)]{jackson2012}
Jackson, A. P. \& Wyatt, M. C. 2012. Debris from terrestrial planet formation: the Moon-forming collision. {\it MNRAS}, 425, 657.

\bibitem[Jacobson et al.(2014)]{jacobson14}
{Jacobson, S. A., Morbidelli, A.,  Raymond, S. N., O'Brien, D. P., Walsh, K. J., \& Rubie, D. C.} 2014. Highly siderophile elements in Earth's mantle as a clock for the Moon-forming impact. {\it Nature}, 508, 84.

\bibitem[Jacquet et al.(2011)]{jacquet11}
{{Jacquet}, E., {Balbus}, S., \& {Latter}, H.} 2011. On linear dust-gas streaming instabilities in protoplanetary discs. {\it \mnras}, 415, 3591


\bibitem[Jewitt \& Luu(1993)]{jewitt1993}
Jewitt, D., \& Luu, J. 1993. Discovery of the candidate Kuiper Belt object 1992 QB1. {\it Nature}, 362, 730.


\bibitem[Johansen \& Youdin(2007)]{jy07}
{{Johansen}, A. and {Youdin}, A. N.} 2007. Protoplanetary disk turbulence driven by the streaming instability: nonlinear saturation and particle concentration. {\it \apj}, 662, 627


\bibitem[Johansen et al.(2012)]{JohYou12}
{{Johansen}, A., {Youdin}, A. N., \& {Lithwick}, Y.} 2012. Adding particle collisions to the formation of asteroids and Kuiper belt objects via streaming instabilities. {\it \aap}, 537, A125.

\bibitem[Johansen et al.(2015b)]{johansen15b}
{{Johansen}, A.,  {MacLow}, M.-M., {Lacerda}, P., \& {Bizzarro}, M.} 2015b. Growth of asteroids, planetary embryos, and Kuiper belt objects by chondrule accretion. {\it Sci. Adv.}, 115109.

\bibitem[Johansen et al.(2015a)]{johansen15}
{{Johansen}, A., {Jacquet}, E., {Cuzzi}, J. N., {Morbidelli}, A., \& {Gounelle}, M.} 2015a. New paradigms For asteroid formation.  arXiv:1505.02941.

\bibitem[Johnson \& Melosh (2012)]{johnson2012}
Johnson, B. C., \& Melosh, H. J. 2012. Formation of spherules in impact produced vapor plumes. {\it Icarus}, 217, 416.

\bibitem[Johnson et al.(2012)]{john2012}
Johnson, B. C., Lisse, C. M., Chen, C. H. et al. 2012. A self-consistent model of the circumstellar debris created by a giant hypervelocity mpact in the HD 172555 system. {\it ApJ}, 761, 45.

\bibitem[Kalas et al.(2013)]{kalas2013}
Kalas, P., Graham, J. R., Fitzgerald, M. P., \& Clampin, M. 2013. STIS coronagraphic imaging of Fomalhaut: main belt structure and the orbit of Fomalhaut b. {\it ApJ}, 775, 56.

\bibitem[Kennedy \& Kenyon (2008)]{kennedy2008}
Kennedy, G. M., \& Kenyon, S. J. 2008. Planet formation around stars of various masses: The snow line and the frequency of giant planets. {\it ApJ}, 673, 502.

\bibitem[Kennedy \& Wyatt (2014)]{kennedy2014}
Kennedy, G. M., \& Wyatt, M. C. 2014. Do two-temperature debris discs have multiple belts?. {\it MNRAS}, 444, 3164.

\bibitem[Kenyon \& Bromley (2004)]{kenyon04}
{{Kenyon}, S. J. \& {Bromley}, B. C.} 2004. Detecting the dusty debris of terrestrial planet formation. {\it ApJL}, 602, 133.


\bibitem[Kenyon \& Bromley (2006)]{kb06}
{{Kenyon}, S. J. \& {Bromley}, B. C.} 2006. Prospects for detection of catastrophic collisions in debris disks. {\it \aj}, 131, 1837.

\bibitem[Kenyon \& Bromley (2012)]{kenyon12}
{{Kenyon}, S. J. \& {Bromley}, B. C.} 2012. Coagulation calculations of icy planet formation at 15-150 AU: a correlation between the maximum radius and the slope of the size distribution for trans-Neptunian objects. {\it \aj}, 143, 63.


\bibitem[Kleine et al.(2009)]{kleine09}
{{Kleine}, T.,  {Touboul}, M., {Bourdon}, B., et al.} 2009. Hf-W chronology of the accretion and early evolution of asteroids and terrestrial planets. {\it {\gca}}, 73, 5150.

\bibitem[Kokubo \& Ida (1998)]{ki98}
{{Kokubo}, E. \& {Ida}, S.} 1998. Oligarchic growth of protoplanets. {\it Icarus}, 131, 171.

\bibitem[Kral et al.(2015)]{kral2015}
Kral, Q., Th\'ebault, P., Augereau, J.-C., Boccaletti, A., \& Charnoz, S. 2015. Oligarchic growth of protoplanets. {\it A\&A}, 573, 39.

\bibitem[Kretke \& Lin (2007)]{kretke2007}
Kretke, K. A. \& Lin, D. N. C. 2007. Grain retention and formation of planetesimals near the snow line in MRI-driven turbulent protoplanetary disks. {\it ApJL}, 664, 55.








\bibitem[Liou \& Zook (1999)]{liou1999}
Liou, J.-C., \& Zook, H. A. 1999. Signatures of the giant planets imprinted on the Edgeworth-Kuiper Belt dust disk. {\it AJ}, 118, 580.


\bibitem[Lisse et al.(2009)]{lisse2009}
Lisse, C. M., Chen, C. H., Wyatt, M. C. et al. 2009. Abundant circumstellar silica dust and SiO gas created by a giant hypervelocity collision in the ~12 Myr HD172555 system. {\it ApJ}, 701, 2019.


\bibitem[Lyra \& Lin (2013)]{lyra13}
{{Lyra}, W. \& {Lin}, M.-K.} 2013. Steady state dust distributions in disk vortices: observational predictions and applications to transitional disks. {\it \apj}, 775, 17.

\bibitem[Malhotra (2015)]{malhotra2015}
Malhotra, R. 2015. The mass distribution function of planets. {\it ApJ}, 808, 71.

\bibitem[Mann et al.(2007)]{mann2007}
Mann, I., Murad, E., \& Czechowski, A. 2007. Nanoparticles in the inner Solar System. {\it  Plan. Spc. Sci}, 55, 1000.



\bibitem[Melis et al.(2012)]{melis2012}
Melis, C., Zuckerman, B., Rhee, J. H. et al. 2012. Rapid disappearance of a warm, dusty circumstellar disk. {\it Nature}, 487, 74.

\bibitem[Meng et al.(2012)]{meng2012}
Meng, H. Y. A., Rieke, G. H., Su, K. Y. L. et al. 2012. Variability of the infrared excess of extreme debris disks. {\it ApJL}, 751, 17.

\bibitem[Meng et al.(2014)]{meng2014}
Meng, H. Y. A., Su, K. Y. L., Rieke, G. H. 2014. Large impacts around a solar-analog star in the era of terrestrial planet formation. {\it Science}, 345, 1032.


\bibitem[Meng et al.(2015)]{meng2015}
Meng, H. Y. A., Su, K. Y. L., Rieke, G. H. et al. 2015. Planetary collisions outside the Solar System: time domain characterization of extreme debris disks. {\it ApJ}, 805, 77.

\bibitem[Mennesson et al.(2014)]{mennesson2014}
Mennesson, B., Millan-Gabet, R., Serabyn, E. et al. 2014. Constraining the exozodiacal luminosity function of main-sequence stars: complete results from the Keck Nuller mid-infrared surveys. {\it ApJ}, 797, 119.

\bibitem[Minato et al.(2006)]{minato2006}
Minato, T., K\"ohler, M., Kimura, H., Mann, I., \& Yamamoto, T. 2006. Momentum transfer to fluffy dust aggregates from stellar winds. {\it A\&A}, 452, 701.

\bibitem[Mittal et al.(2015)]{mittal2015}
Mittal, T., Chen, C. H., Jang-Condell, H. et al. 2015. The Spitzer Infrared Spectrograph Debris Disk Catalog. II. silicate feature analysis of unresolved targets. {\it ApJ}, 798, 87.



\bibitem[{{Morbidelli} {et~al.}(2015a){Morbidelli}, {Walsh},
  {O'Brien}, {Minton}, \& {Bottke}}]{morbidelli15}
{Morbidelli}, A., {Walsh}, K. J., {O'Brien}, D. P., {Minton}, D. A., \&
  {Bottke}, W.F. 2015a. The dynamical evolution of the asteroid belt. arXiv:1501.06204.

\bibitem[Morbidelli et al.(2015b)]{morbidelli15b}
{{Morbidelli}, A., {Lambrechts}, M., {Jacobson}, S., \& {Bitsch}, B.} 2015b. The great dichotomy of the Solar System: Small terrestrial embryos and massive giant planet cores. {\it \icarus}, 258, 418.


\bibitem[Mulders et al. (2015)]{mulders15}
Mulders, G. D., Pascucci, I., \& Apai, D. 2015. An Increase in the Mass of Planetary Systems around Lower-mass Stars. {\it \apj}, 814, 130


\bibitem[Nesvorn\'y et al.(2003)]{nesvorny2003}
Nesvorn\'y, D., Bottke, W. F., Levison, H. F., \& Dones, L. 2003. Recent origin of the Solar System dust bands. {\it ApJ}, 591, 486.

\bibitem[Nesvorn\'y et al.(2010)]{nesvorny2010}
Nesvorn\'y, D., Jenniskens, P., Levison, H. A., Bottke, W. F., Vokrouhlick\'y, D., \& Gounelle, M. 2010. Cometary origin of the zodiacal cloud and carbonaceous micrometeorites. implications for hot debris disks. {\it ApJ}, 713, 816.

\bibitem[{{O'Brien} \& {Greenberg}(2003)}]{obrien03}
{O'Brien}, D. P., \& {Greenberg}, R. 2003. Steady-state size distributions for collisional populations:. analytical solution with size-dependent strength. {\it \icarus}, 164, 334.

\bibitem[O'Brien \& Sykes(2011)]{obrien2011}
O'Brien, D. P., \& Sykes, M. V. 2011. The origin and evolution of the asteroid belt—implications for Vesta and Ceres. {\it Spc. Sci. Rev.}, 163, 11.

\bibitem[Ootsubo et al.(2009)]{ootsubo2009}
Ootsubo, T., Ueno, M., Ishiguro, M. et al. 2009. Mid-Infrared spectrum of the zodiacal light observed with AKARI/IRC. {\it  ASP Conf. Ser.}, 418, 395.

\bibitem[Ormel \& Klahr (2010)]{ormel10}
{{Ormel}, C.W. and {Klahr}, H.H.} 2010, {The effect of gas drag on the growth of protoplanets.}, {\it \aap}, 520A, 43.

\bibitem[Pan et al.(2011)]{pan11}
{{Pan}, L., {Padoan}, P., {Scalo}, J., {Kritsuk}, A. G., \& {Norman}, M. L.} 2011. Turbulent clustering of protoplanetary dust and planetesimal formation. {\it \apj}, 740, 6.

\bibitem[Papoular(1983)]{papoular1983}
Papoular, R., \& P\'egouri\'e, B. 1983. The IR silicate features as a measure of grain size in circumstellar dust. {\it A\&A}, 128, 335.




\bibitem[Plavchan et al.(2005)]{plavchan2005}
Plavchan, P., Jura, M., \& Lipscy, S. J. 2005. Where are the M dwarf disks older than 10 million years?. {\it ApJ}, 631, 1161.



\bibitem[Rieke et al.(2015)]{rieke2015}
Rieke, G. H., G\'asp\'ar, A., \& Ballering, N. P. 2015. Magnetic grain trapping and the hot excesses around early-type stars. arXiv:1511.04998, {\it ApJ, accepted}.

\bibitem[Roberge et al.(2012)]{roberge2012}
Roberge, Aki, Chen, C. H., Millan-Gabet, R. et al. 2012. The exozodiacal dust problem for direct observations of exo-earths. {\it PASP}, 124, 799.



\bibitem[Schneider et al. (2014)]{schneider14}
Schneider, G., Grady, C. A., Hines, D. C., et al. 2014. Probing for exoplanets hiding in dusty debris disks: disk imaging, characterization, and exploration with HST/STIS multi-roll coronagraphy. {\it AJ}, 148, 59

\bibitem[Shariff \& Cuzzi(2011)]{ShaCuz11}
{{Shariff}, K. and {Cuzzi}, J. N.} 2011. Gravitational instability of solids assisted by gas drag: slowing by turbulent mass diffusivity. {\it \apj}, 738, 73.

\bibitem[Sheehan \& Eisner(2014)]{sheehan2014}
Sheehan, P. D., \& Eisner, J. A. 2014. Constraining the disk masses of the class I binary protostar GV Tau. {\it ApJ}, 791, 19.

\bibitem[Sierchio et al.(2014)]{sierchio2014}
Sierchio, J. M., Rieke, G. H., Su, K. Y. L., \& G\'asp\'ar, A. 2014. The decay of debris disks around Solar-type stars. {\it ApJ}, 785, 33.

\bibitem[{{Simon} {et~al.}(2015){Simon}, {Armitage}, {Li}, \&
  {Youdin}}]{simon15}
{Simon}, J. B., {Armitage}, P. J., {Li}, R., \& {Youdin, A. N.} The initial mass and size distribution of planetesimals. I. the effect of resolution, gravity, and initial conditions in streaming instability calculations.  2015,  arXiv:1512.00009.
  
\bibitem[Smith \& Terrile(1984)]{smith1984}
Smith, B. A., \& Terrile, R. J. 1984. A circumstellar disk around Beta Pictoris. {\it Science}, 226, 1421.

\bibitem[Soderblom et al.(2014)]{soderblom2014}
Soderblom, D. R., Hillenbrand, L. A., Jeffries, R. D., Mamaje, E. E., \& Naylor, T. 2014. Ages of young stars. in {\it Protostars \& Planets VI}, ed. H. Beuther, R.S. Klessen, C.P. Dullemond, \& T. Henning, UA Press: Tucson, pp 219-241.

\bibitem[Soummer et al. (2014)]{soummer14}
Soummer, R., Perrin, M. D., Pueyo, L. et al. 2014. Five debris disks newly revealed in scattered light from the Hubble Space Telescope NICMOS archive. {\it ApJL}, 786, 23.


\bibitem[Stapelfeldt et al.(2004)]{stapelfeldt2004}
Stapelfeldt, K. R., Holmes, E. K., Chen, C. H. et al. 2004. First Look at the Fomalhaut debris disk with the Spitzer Space Telescope. {\it  \apjs}, 154, 458.

\bibitem[Stock et al.(2010)]{stock2010}
Stock, N. D., Su, K. Y.L., Liu, W. et al. 2010. The structure of the $\beta$ Leonis debris disk. {\it ApJ}, 724, 1238.

\bibitem[Su et al.(2013)]{su2013}
Su, K. Y. L., Rieke, G. H., Malhotra, R., et al. 2013. Asteroid belts in debris disk twins: Vega and Fomalhaut. {\it ApJ}, 763, 118.

\bibitem[Su \& Rieke(2014)]{su2014}
Su, K. Y. L., \& Rieke, G. H. 2014. Signposts of multiple planets in debris disks. {\it IAU Symp.}, 299, 318.

\bibitem[Szalay et al.(2015)]{szalay2015}
Szalay, J., Piquette, M., \& Horanyi, M. 2015. Dust measurements by the Student Dust Counter onboard the New Horizons Mission to Pluto. {\it Lun. Plan. Inst.}, 1832, 1701.

\bibitem[{{Takahashi} \& {Inutsuka}}(2014)]{takahashi14}
{{Takahashi}, S. Z. and {Inutsuka}, S.-I.} 2014. Two-component secular gravitational instability in a protoplanetary disk: a possible mechanism for creating ring-like structures. {\it \apj}, 794, 55.

\bibitem[Th\'ebault \& Augereau(2007)]{thebault2007}
Th\'ebault, P. \& Augereau, J.-C. 2007. Collisional processes and size distribution in spatially extended debris discs. {\it A\&A}, 472, 169.

\bibitem[van Boekel et al.(2004)]{vanboekel2004}
van Boekel, R., Min, M., Leinert, Ch, et al. 2004. The building blocks of planets within the `terrestrial' region of protoplanetary disks. {\it Nature}, 432, 479.


\bibitem[Vitense et al.(2012)]{vitense2012}
Vitense, Ch, Krivov, A. V., Kobayashi, H., \& L\"ohne, T. 2012. An improved model of the Edgeworth-Kuiper debris disk. {\it A\&A}, 540, 30.


\bibitem[Walsh {et~al.}(2011)]{walsh11}
{Walsh}, K. J., {Morbidelli}, A., {Raymond}, S. N., {O'Brien}, D. P., \&
  {Mandell}, A. M. 2011. A low mass for Mars from Jupiter's early gas-driven migration. {\it \nat}, 475, 206.

\bibitem[Walsh {et~al.}(2012)]{walsh12}
{Walsh}, K. J., {Morbidelli}, A., {Raymond}, S. N., {O'Brien}, D. P., \&
  {Mandell}, A. M. 2012. Populating the asteroid belt from two parent source regions due to the migration of giant planets ``The Grand Tack". {\it Meteor. \& Plan. Sci.}, 47, 1941.



\bibitem[Weidenschilling (1980)]{stu80}
{{Weidenschilling}, S.J.} 1980,  {Dust to planetesimals - Settling and coagulation in the solar nebula.} {\it Icarus}, 44, 172

\bibitem[Weissman(1984)]{weissman1984}
Weissman, P. R. 1984. The VEGA particulate shell - comets or asteroids? {\it Science}, 224, 987.


\bibitem[{Whipple}(1972)]{whi72}
{{Whipple}, F. L.} 1972. On certain aerodynamic processes for asteroids and comets. in {\it From Plasma to Planet}, Ed. {{Elvius}, A.}, Wiley: New York, p 211.


\bibitem[{Williams \& Cieza}(2011)]{williams11}
{{Williams}, J. P. and {Cieza}, L. A.} 2011. Protoplanetary disks and their evolution. {\it \araa}, 49, 67.



\bibitem[{Wyatt et al.}(2007)]{wyatt2007}
Wyatt, M. C., Smith, R., Su, K. Y. L. et al. 2007. Steady State evolution of debris disks around A stars. {\it ApJ}, 663, 365.

\bibitem[{Wyatt}(2008)]{wyatt2008}
Wyatt, M. C. 2008. Evolution of debris disks. {\it \araa}, 46, 339.


\bibitem[Youdin(2010)]{houches10}
{{Youdin}, A. N.} 2010. From grains to planetesimals, ESA, Ed. {{T. Montmerle, D.~Ehrenreich, \& A.-M.~Lagrange}}, 41, 187.

\bibitem[{Youdin}(2011)]{you11a}
---. 2011. On the formation of planetesimals via secular gravitational instabilities with turbulent stirring, {\it \apj}, 731, 99.

\bibitem[{{Youdin}(2011{\natexlab{b}})}]{you11b}
---. 2011{\natexlab{b}}, {The Exoplanet Census: A General Method Applied to Kepler}, {\it \apj}, 742, 38

\bibitem[{{Youdin} \& {Chiang}(2004)}]{yc04}
{Youdin}, A. N., \& {Chiang}, E. I. 2004. Particle pileups and planetesimal formation. {\it \apj}, 601, 1109.

\bibitem[Youdin \& Goodman(2005)]{yg05}
{{Youdin}, A. N. and {Goodman}, J.} 2005. Streaming instabilities in protoplanetary disks. {\it \apj}, 620, 459.

\bibitem[Youdin \& Johansen(2007)]{yj07}
{{Youdin}, A. N. \& {Johansen}, A.} 2007. Protoplanetary disk turbulence driven by the streaming instability: linear evolution and numerical methods. {\it \apj}, 662.

\bibitem[{{Youdin} \& {Kenyon}(2012)}]{YouKen12}
{Youdin}, A. N., \& {Kenyon}, S. J. 2012. From disks to planets. in {\it Planets, Stars and Stellar Systems} ed. T. D. Oswalt, L. M. French, P. Kalas, Springer: Dordrecht, pp 1-62.

\bibitem[{{Youdin} \& {Lithwick}(2007)}]{yl07}
{Youdin}, A. N., \& {Lithwick}, Y. 2007. Particle stirring in turbulent gas disks: Including orbital oscillations. {\it Icarus}, 192, 588.

\bibitem[Youdin \& Shu(2002)]{ys02}
{{Youdin}, A. N. \& {Shu}, F. H.} 2002. Planetesimal formation by gravitational instability. {\it \apj}, 580, 494.

\bibitem[Zsom et al.(2010)]{zsom10}
{{Zsom}, A., {Ormel}, C .W., {G{\"u}ttler}, C.,  {Blum}, J. \& {Dullemond}, C. P.} 2010. The outcome of protoplanetary dust growth: pebbles, boulders, or planetesimals? II. Introducing the bouncing barrier. {\it \aap}, 513A, 57.





\end{thebibliography}
\end{document}